\pdfoutput=1

\documentclass[11pt]{article}

\usepackage{acl}

\usepackage{times}
\usepackage{latexsym}
\usepackage{amssymb}
\usepackage{multirow}
\usepackage{booktabs}
\usepackage{graphicx}

\usepackage[T1]{fontenc}

\usepackage[utf8]{inputenc}
\usepackage{enumitem}

\usepackage{microtype}
\usepackage{pifont}
\usepackage{amsmath}
\newcommand{\cmark}{\ding{51}}%
\newcommand{\xmark}{\ding{55}}%
\usepackage{color, colortbl}
\definecolor{Gray}{gray}{0.9}

\title{Transferable Embedding Inversion Attack:\\ Uncovering Privacy Risks in Text Embeddings without Model Queries
}

\author{Yu-Hsiang Huang \\
  National Taiwan University \\
  {r11922053@csie.ntu.edu.tw} \\\And
  Yu-Che Tsai \\
  National Taiwan University \\
  {f09922081@csie.ntu.edu.tw} \\ \And
  Hsiang Hsiao \\
  National Taiwan University \\
  {r12946003@ntu.edu.tw} \\ \AND
  Hong-Yi Lin \\
  National Taiwan University \\
  {b09902100@csie.ntu.edu.tw} \\ \And
  Shou-De Lin \\
  National Taiwan University \\
  {sdlin@csie.ntu.edu.tw} \\
}

\begin{document}

\maketitle
\begin{abstract}
This study investigates the privacy risks associated with text embeddings, focusing on the scenario where attackers cannot access the original embedding model. Contrary to previous research requiring direct model access, we explore a more realistic threat model by developing a transfer attack method. This approach uses a surrogate model to mimic the victim model's behavior, allowing the attacker to infer sensitive information from text embeddings without direct access. Our experiments across various embedding models and a clinical dataset demonstrate that our transfer attack significantly outperforms traditional methods, revealing the potential privacy vulnerabilities in embedding technologies and emphasizing the need for enhanced security measures.
\end{abstract}

\section{Introduction}

Text embeddings serve as universal representations of textual data, which can be utilized as features for various downstream tasks. Recent developments in text embedding models~\cite{sentence-T5,sentenceBERT} have significantly streamlined the process of generating embeddings. Additionally, systems that employ large language models (LLMs) often incorporate a vector database of text embeddings to store and infuse domain specific knowledge or auxiliary data. Retrieval-augmented generation (RAG)~\cite{RAG} is a typical example that enhances LLMs' knowledge by incorporating retrieved documents into the model's prompt. This has led to a growing adoption of vector database services like Chroma\footnote{\url{https://docs.trychroma.com/}} and Faiss~\cite{Fassis}, known for their efficient and scalable embedding searches. In these databases, only the text embeddings are shared with third-party services, not the actual text, leading these platforms to claim that storing embeddings is safe and encouraging the upload of private data.

Despite the bright sight of text embeddings and vectors, a natural question arises: Does sending text embedding to an online service really expose zero privacy risk given that the original text might contain sensitive information? To answer this question, researchers started to investigate the \textit{embedding inversion attack}, which aims to reconstruct the input data from its embedding. 
In the image domain, prior works~\cite{CV_embedding_inversion,dosovitskiy2016inverting} on computer vision demonstrated that it is possible to reproduce the input image from their visual embeddings. In the text domain, the pioneering work~\cite{information_leak} attempts to infer a bag of words from embeddings. Along similar lines, the following research~\cite{morris2023text} further reveals that an adversary can recover 92\% of a 32-token text input given embeddings from a T5-based pre-trained transformer.

Although existing works~\cite{information_leak, morris2023text, GEIA} have studied the privacy risks of text embeddings, the observed threats essentially rely on a strong assumption, which is that the adversary has query access to the embedding model. By querying the embedding model extensively, the adversary can either iteratively edit the input text such that the text is as close as possible to a given embedding or obtain a large amount of paired data to reverse-engineer the embedding model accordingly. Here, we argue that such privileged knowledge might not always be available in real-world scenarios. For instance, consider the data leakage of an online vector database where only a small number of documents and their associated text embeddings were exposed to the adversary. In that case, the adversary was passively offered a small number of query pairs, while querying the embedding model is not allowed. Inspired by this, this work particularly focuses on the privacy risk without assuming the accessibility of the original embedding model for querying; instead, only a small portion of paired document-embedding data is available.

We consider the problem of a black-box attack, where the target victim model is entirely hidden from the attacker. In this setting, standard white-box attacks~\cite{kugler2021invbert} or even query-based black-box attacks~\cite{GEIA, morris2023text} become ineffective. 
As the victim model becomes invisible to the adversary, we present an alternative solution to attack the victim model through a transfer attack. 
The transferability property of an attack is satisfied when an attack developed for a particular machine learning model (i.e., a surrogate model) is also effective against the target model.
Specifically, our transfer attack aims to achieve two goals: 1) \textbf{Encoder stealing} attempts to learn a \textit{surrogate model} to steal the victim model only through their returned representations. If the surrogate model successfully replicates the victim model, the adversary gains query access to some extent. 
2) \textbf{Threat model transferability} enables the adversary to build a threat model by attacking the surrogate model and hopes the threat model can also successfully fool the victim black-box model.\footnote{Code is publicly available at \url{https://github.com/coffree0123/TEIA}}

To achieve the first goal, an off-the-shelf text embedding model (e.g., GTR-T5~\cite{GTR}) followed by a MLP-based adapter is used as the surrogate model. The surrogate model is then optimized with our proposed consistency regularization loss to mimic the behavior of the victim model. 
To achieve the second goal, we use the adversarial training to mitigate the embedding discrepancy between the surrogate and victim models and thus improve the attack transferability. 

To validate the effectiveness of our attack, we perform extensive experiments on 3 popular embedding models, including Sentence-BERT~\cite{sentenceBERT}, Sentence-T5~\cite{sentence-T5}, and OpenAI text embedding. Experimental results show that the transfer attack can be 40\%-50\% more effective than the standard attack approach. The key factors for stealing the victim model are discussed in Sec.~\ref{sec:surrogate-analysis}. To study the privacy risk on a specific threat domain, we conduct a case study on the MIMIC-III clinical note dataset. Results demonstrated in Sec.~\ref{sec:case-study} show that our transfer attack can identify sensitive attributes (e.g., age, sex, disease) with 80\%-99\% accuracy. 

\section{Preliminary}

\subsection{Embedding inversion attack}
Given a sequence of text tokens $x \in V^{n}$, the text encoder $\phi: V^{n} \rightarrow \mathbb{R}^{d}$ will map the text $x$ into a fixed-length vector $\phi(x) \in \mathbb{R}^{d}$ which is the text embedding. 
An embedding inversion attack is a specific type of embedding attack that aims to reconstruct the original text $x$ from its text embedding $\phi(x)$. Specifically, the attacker seeks to find a function $f$ to approximate the inversion function of $\phi$ as:
\begin{equation}
    f(\phi(x)) \approx \phi^{-1}(\phi(x)) = x.
\end{equation}
According to the attack target, the embedding inversion attack can be categorized into: (i) token-level inversion~\cite{privacy_risks,information_leak} and (ii) sentence-level inversion~\cite{GEIA,morris2023text}. As inverting the whole sentence could potentially reveal more privacy risks, we focus on recovering the whole sentence from its text embedding in this work.

\noindent \textbf{Base attack model.}
\label{sec:embedding-inversion}
To reconstruct the original text sequence $x = w_0w_1...w_{u}$ from its corresponding text embedding $\phi(x)$, a recent work~\cite{GEIA} proposed the attack model as a generative task.
This involves minimizing the standard language model loss with teacher forcing \cite{TeachingForcing}. This loss function is defined as:
\begin{equation}
\label{eq:GEIA-loss}
\mathcal{L}_{LM}=-\sum_{i=1}^{u}\log( Pr(w_i | \phi(x), w_0, \ldots, w_{i-1})    
\end{equation}
\begin{figure*}[t!]
  \centering
  \includegraphics[width=\linewidth]{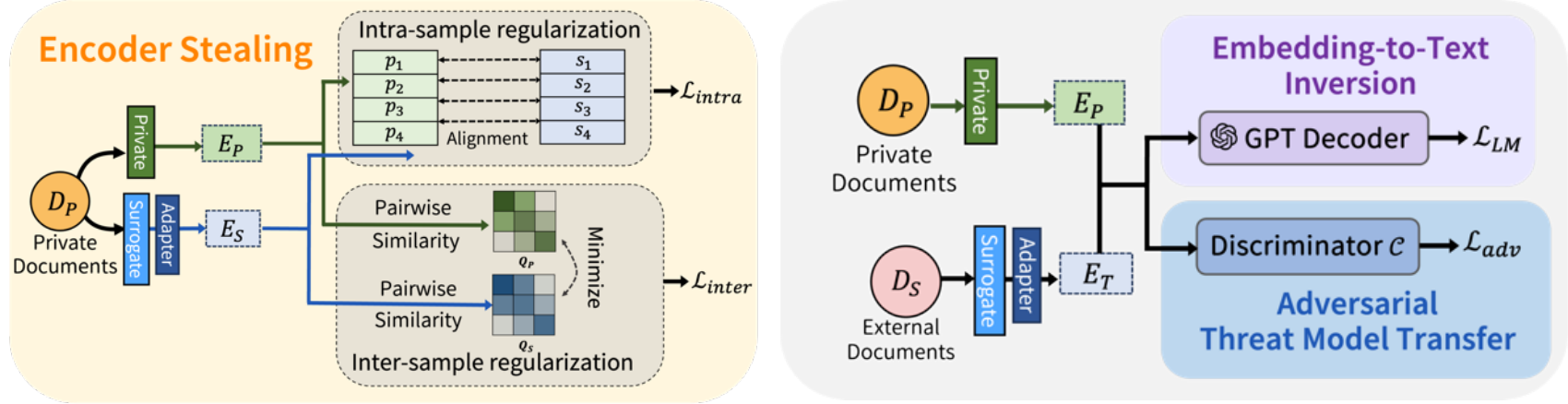}
  \caption{Model architecture of the transferable embedding inversion attack.}
  \label{fig:model-architecture}
\end{figure*}

\subsection{Transferable embedding inversion attack}
\textbf{Motivation.} 
In practical scenarios, attackers might access text embeddings without the ability to query the generating model directly. For instance, a data breach might expose embeddings from a health chatbot containing encoded patient information or from a job recommendation platform with details on resumes and job listings. These situations demonstrate the risk of sensitive data exposure even when direct interaction with the embedding model is not possible and motivate our research into developing methods to study privacy under such constrained conditions.

\noindent \textbf{Attacker's goals:} The attacker aims to achieve the following two goals:
\vspace{-\topsep}
\begin{itemize}[leftmargin=*]
    \item \textbf{\textit{Goal 1 (Stealing Text Encoder)}}: The attacker seeks to find a surrogate encoder $\hat{\phi}$ to steal the anonymous embedding model $\phi$. In particular, we expect an optimal surrogate model that satisfies:
    \begin{equation}
        \hat{\phi}(x) \approx \phi(x) ;~ \forall x \in \mathcal{X},
    \end{equation}
    where $\hat{\phi}$ gives the similar output embedding as $\phi$ and $\mathcal{X}$ denotes the domain of an input text $x$.
    \item  \textbf{\textit{Goal 2 (Threat Model Transferability)}}: Given the surrogate model $\hat{\phi}$, the attacker constructs the surrogate dataset $D_S = \{(x, \hat{\phi}(x))\}$ which consists of pairs of documents and their text embeddings. The threat model $\mathcal{T}: \hat{\phi}(x) \rightarrow x$ is then built to attack $\hat{\phi}$ using $D_S$. Finally, $\mathcal{T}$ is used to perform a transfer attack on $\phi$.
\end{itemize}

\noindent \textbf{Attacker's background knowledge.}
To clarify the scope of the attacker's background knowledge, we make the following assumptions:
\vspace{-\topsep}
\begin{itemize}[leftmargin=*]
    \item \textbf{\textit{Assumption 1 (Anonymous Embedding Model)}}: Our attack follows the realm of black-box attack, where the attacker is not aware of the model weights or the architecture of $\phi$. Unlike prior works that presume query access to $\phi$, we further eliminate such knowledge and make $\phi$ completely hidden from the attacker.

    \item \textbf{\textit{Assumption 2 (Leaked Dataset)}}: We assume a dataset $D_{L} = \{ (x, \phi(x)) \}$ is exposed to the attacker due to potential data leakage from an online vector database or embedding services. In a practical sense, we consider $D_L$ to be a small dataset.
\end{itemize}


\section{Methodology}
As illustrated in Figure~\ref{fig:model-architecture}, our transfer attack pipeline consists of three major components. First, the encoder stealing aims to learn a surrogate model to mimic the behavior of the victim model $\phi$. This goal is achieved by optimizing the surrogate model with our intra- and inter-consistency regularization losses. Second, we adopt adversarial training to make the surrogate embedding indistinguishable from the private embedding and improve attack transferability. Finally, embedding-to-text leverages a GPT-based decoder to invert embeddings to their original text sequence.

\subsection{Encoder Stealing with a Surrogate Model}
\textbf{The surrogate model.}
The primary objective of the surrogate model $\hat{\phi}$ is to steal the black-box embedding model $\phi$ through the leaked dataset $D_L$. The surrogate model consists of two components: a surrogate encoder and an adapter. The surrogate encoder is a pre-trained text embedding model used to generate the initial embedding of input text $x$. A simple linear transformation is used as the adapter to convert the initial embedding such that the resulting representation could be aligned with $\phi(x)$. 
Adding an adapter behind the surrogate encoder has two advantages: (1) We do not need to fine-tune the surrogate encoder during training; only the adapter's model weight needs to be adjusted. (2) The adapter can solve the issue of the output dimension of the surrogate encoder being inconsistent with $\phi(x)$.


\noindent\textbf{Optimizing the surrogate model with consistency regularization.} 
To achieve Goal 1, we design two types of regularization terms to enforce $\hat{\phi}$ acts similarly to $\phi$. Given a batch of $N$ samples from $D_L$, we have $\mathbf{E}_P = \phi(x)$ and $\mathbf{E}_S=\hat{\phi}(x)$, where $\mathbf{E}_P \in \mathbb{R}^{N\times d}$ and $\mathbf{E}_S \in \mathbb{R}^{N\times d}$ denote the private and surrogate embedding matrix, respectively.
Inspired by the concept of stealing image encoder~\cite{liu2022stolenencoder}, the intra-consistency regularization aims to minimize the distance between the $\mathbf{E}_P$ and $\mathbf{E}_S$, which is described using the following loss:
\begin{equation}
    \mathcal{L}_{intra}(\mathbf{E}_P, \mathbf{E}_S)= MSE(\mathbf{E}_P, \mathbf{E}_S).
\end{equation}
Here, we use the mean squared error to measure the distance between two matrices. $\mathcal{L}_{intra}$ is small if $\hat{\phi}$ and $\phi$ produce similar feature vectors for an input.

However, simply optimizing $\mathcal{L}_{intra}$ only considers point-wise information and ignores pairwise semantic information between documents. Therefore, we designed an additional inter-consistency regularization term to enable our surrogate model to simultaneously preserve the relative semantic relationship between documents. Specifically, we first calculate the in-batch pairwise cosine similarity matrix $\mathbf{Q}_P \in \mathbb{R}^{N\times N}$ from the private embedding $\mathbf{E}_P$ as:
\begin{equation}
    \label{eq:pairwise}
    \mathbf{Q}_P = \Tilde{\mathbf{Q}}_P \Tilde{\mathbf{Q}}_P^{\top}; \ \Tilde{\mathbf{Q}}_{P[i,:]} = \mathbf{E}_{P[i,:]} / \lVert \mathbf{E}_{P[i,:]} \rVert_2.
\end{equation}
Similarly, the pairwise similarity $\mathbf{Q}_S$ could be obtained from $\mathbf{E}_S$ using E.q.~\ref{eq:pairwise}. Finally, we define the similarity-preserving regularization loss as: 
\begin{equation}
    \mathcal{L}_{inter}(\mathbf{Q}_P, \mathbf{Q}_S) = \frac{1}{N^2} \lVert \mathbf{Q}_P - \mathbf{Q}_S  \rVert^2_F,
\end{equation}
where $\lVert \cdot \rVert_F$ is the Frobenius norm of a matrix. 
We let $\mathcal{L}_{surrogate} = \mathcal{L}_{intra}+\mathcal{L}_{inter}$ be the objective loss function for stealing the text encoder with $\hat{\phi}$.

\subsection{Adversarial Threat Model Transferability}
To achieve Goal 2, the attacker leverages an external corpus and utilizes the surrogate model to create the surrogate dataset $D_S = \{ (x, \hat{\phi}(x) \}$.
As the GPT decoder is primarily trained on $D_S$ using the surrogate-generated embeddings, a key obstacle here is the difference in representation between the surrogate and private embeddings, which can hinder the effectiveness of the attack when applied to the latter. To overcome this, we employ adversarial training techniques~\cite{DANN}. This method involves training a discriminator $\mathcal{C}$ to distinguish between the surrogate embedding $\mathbf{E}_T$ and private embedding $\mathbf{E}_P$ while simultaneously optimizing $\hat{\phi}$ to generate embeddings that the discriminator cannot differentiate. Note that here we denote $\mathbf{E}_T$ as the surrogate embeddings generated from external documents. Formally, the adversarial training is described as:
\begin{equation}
\begin{aligned}
    \mathcal{L}_{adv} =  \underset{\hat{\phi}}{\min} \; \underset{\mathcal{C}}{\max} \; &~\mathbb{E}_{e_p \sim \mathbf{E}_P}[\log{\mathcal{C}(e_p)}] + \\ &~\mathbb{E}_{e_t \sim \mathbf{E}_{T}}[\log{(1 - \mathcal{C}(e_t))}],
    \label{eq:domain-adv}
\end{aligned}
\end{equation}
where $\log \mathcal{C}(e_p)$ and $\log \mathcal{C}(e_t)$ represent the expected value of the logarithmic probability of the domain classifier $\mathcal{C}$.
During the training phase, we utilize a sequential training strategy, alternating focus between the discriminator and the surrogate encoder. 
\subsection{Training Pipeline}
In every training iteration, we sample a batch data from both $D_L$ and $D_S$. The leaked dataset is used for encoder stealing and embedding-to-text training.
Considering the leaked dataset $D_L$ could be small, we also apply data augmentation to create more examples. The analysis of data augmentation is studied in Appendix~\ref{appendix:augmentation}.
On the other hand, the surrogate dataset $D_S$ is used for adversarial and embedding-to-text training. 
Finally, we jointly optimize the surrogate model and the base attack model mentioned in Sec.~\ref{sec:embedding-inversion} with the following objective function: $\mathcal{L}_{final} = \mathcal{L}_{LM} + \mathcal{L}_{surrogate} + \mathcal{L}_{adv}.$
\section{Experiment Setup}

\begin{table*}[t!]
\centering
\caption{Comparison of same domain embedding inversion performance between direct and transfer attack. The evaluation is done on QNLI, IMDB, and AGNEWS datasets with embedding models including OpenAI \texttt{text-embeddings-ada-002}, SBERT~\cite{sentenceBERT} and ST5~\cite{sentence-T5}. Higher scores are better for all metrics except PPL.}
\resizebox{\textwidth}{!}{
\begin{tabular}{@{}lcccccccccccc@{}}
\toprule
\multirow{2}{*}{Dataset / Method} & \multicolumn{4}{c}{OpenAI} & \multicolumn{4}{c}{SBERT} & \multicolumn{4}{c}{ST5} \\
\cmidrule(lr){2-5} \cmidrule(lr){6-9} \cmidrule(lr){10-13}
& RougeL & PPL & Cos & LLM-Eval & RougeL & PPL & Cos & LLM-Eval & RougeL & PPL & Cos & LLM-Eval \\
\midrule
\textbf{QNLI} \\
\hspace{1em}Direct Attack & 0.1433 & 40.822 & 0.2797 & 0.2984 & 0.1264 & 27.127 & 0.3257 & 0.3194 & 0.1463 & 42.911 & 0.2226 & 0.2755 \\
\hspace{1em}Transfer Attack & \textbf{0.2226} & \textbf{10.242} & \textbf{0.4772} & \textbf{0.4402} & \textbf{0.1934} & \textbf{11.633} & \textbf{0.4886} & \textbf{0.4280} & \textbf{0.1985} & \textbf{11.808} & \textbf{0.4121} & \textbf{0.3963} \\
\hspace{1em}Improv. (\%) & 55.3\% & 74.9\% & 70.6\% & 47.5\% & 53.0\% & 57.1\% & 50.0\% & 34.0\% & 35.6\% & 72.4\% & 85.1\% & 43.8\% \\
\midrule
\textbf{IMDB} \\
\hspace{1em}Direct Attack & 0.1133 & 20.549 & 0.2692 & 0.3818 & 0.1137 & 34.805 & 0.2891 & 0.3923 & 0.1103 & 24.939 & 0.2678 & 0.3909 \\
\hspace{1em}Transfer Attack & \textbf{0.1991} & \textbf{12.953} & \textbf{0.4297} & \textbf{0.4528} & \textbf{0.1689} & \textbf{14.505} & \textbf{0.4467} & \textbf{0.4475} & \textbf{0.1571} & \textbf{14.839} & \textbf{0.3866} & \textbf{0.4295} \\
\hspace{1em}Improv. (\%) & 75.7\% & 36.9\% & 59.6\% & 18.6\% & 48.5\% & 58.3\% & 54.5\% & 14.0\% & 42.4\% & 40.4\% & 44.3\% & 9.8\% \\
\midrule
\textbf{AGNEWS} \\
\hspace{1em}Direct Attack & 0.0612 & 66.383 & 0.1162 & 0.2979 & 0.0538 & 286.16 & 0.1317 & 0.2742 & 0.0578 & 74.085 & 0.0980 & 0.2905 \\
\hspace{1em}Transfer Attack & \textbf{0.1271} & \textbf{31.159} & \textbf{0.4301} & \textbf{0.4057} & \textbf{0.1067} & \textbf{36.793} & \textbf{0.4110} & \textbf{0.3839} & \textbf{0.1042} & \textbf{40.809} & \textbf{0.3697} & \textbf{0.3706} \\
\hspace{1em}Improv. (\%) & 107.0\% & 53.0\% & 270\% & 36.1\% & 98.3\% & 87.1\% & 212.0\% & 40.0\% & 80.2\% & 44.9\% & 277.2\% & 27.5\% \\
\bottomrule
\end{tabular}
}
\label{tab:main_result_same_domain}
\end{table*}

\noindent \textbf{Victim Embedding Models.}
To assess the embedding inversion attack, we utilize three victim models acting as our blackbox encoders: \texttt{text-embeddings-ada-002} from OpenAI, SBERT~\cite{sentenceBERT}, and ST5~\cite{sentence-T5}. These encoder models remain frozen, with their pre-trained weights employed to generate private embeddings from input text. All encoder models, except for OpenAI, are accessible via Hugging Face.

\noindent \textbf{Datasets.}
Three datasets are used to evaluate the attack performance. 
Qnli~\cite{QNLI} is structured around question-answer pairs and collected from Wikipedia articles. IMDB~\cite{IMDB} comprises movie reviews. AG News~\cite{AGNEWS} includes a diverse collection of news articles. We randomly sample 8000 documents from each dataset to form the leaked dataset $D_L$.
The statistics for these datasets are detailed in Appendix~\ref{appendix:dataset-stats}.

\noindent \textbf{Source Domain of External Dataset.}
Depending on the data domain of the external dataset, the attack scenario can be categorized into in-domain and out-of-domain text reconstruction.
Under the in-domain (out-of-domain) attack setting, we assume the external dataset has the same (different) data domain as the leaked dataset. By default, we employ data from the same domain as the external dataset. 


\noindent \textbf{Competing Method.}
To compare the inversion performance, we employ a generative embedding inversion attack approach~\cite{GEIA} that utilizes the leaked dataset $D_{L}$ and trains the threat model by optimizing E.q.~\ref{eq:GEIA-loss}. Here, this method is referred to as "Direct Attack" to distinguish it from our transfer attack strategy. For a fair comparison, we use the same dialogGPT model~\cite{Dialogpt} as the decoder for both direct and transfer attacks.

\noindent \textbf{Evaluation Metrics.}
We use the following four metrics to evaluate the text reconstruction attack performance. \textbf{RougeL}~\cite{rouge} is used to measure the accuracy and overlap between ground truth and reconstructed text based on n-grams. \textbf{Perplexity}~\cite{ppl} is used to evaluate the performance of language models by measuring how well they predict a given sequence of words.
\textbf{Embedding similarity (Cos)}: To evaluate the semantic similarity in latent space, we utilize Sentence-BERT~\cite{sentenceBERT} to compute the cosine similarity between the ground truth sentences' embedding and the embedding of the generated sentences. \textbf{LLM-Eval}~\cite{LLM_Eval}: We use ChatGPT to provide a score ranging from 0 to 1 to evaluate the relevance between prediction and ground truth. More details can be found in Appendix~\ref{appendix:llm-eval}.

\section{Attack Result}
\label{sec:main-attack}

\subsection{In-domain Text Reconstruction}
Table~\ref{tab:main_result_same_domain} compares the attack performance between direct and transfer attacks on different datasets and victim embedding models. The result shows a significant improvement with more than 40\% in both RougeL and embedding similarity scores when comparing transfer attack to direct attack. It is worth noting that the major difference between direct and transfer attacks is the additional surrogate dataset $D_S$ to enhance the performance of the attack model. Therefore, the improved result indicates a successful transfer of the surrogate model. To better understand the effectiveness of the surrogate model and how well it steals, a detailed discussion can be found in Sec.~\ref{sec:surrogate-analysis}.

\subsection{Out-of-domain Text Reconstruction}
\begin{table}[t!]
\centering
\caption{Comparison of out-of-domain embedding inversion performance between direct and transfer attack.}
\resizebox{\linewidth}{!}{
\begin{tabular}{lcccc}
\toprule
Dataset / Method & RougeL & PPL & Cos & LLM-Eval \\
\midrule
\textbf{QNLI} \\
\hspace{1em}Direct Attack & 0.1264 & 27.127 & 0.3257 & 0.3194 \\
\hspace{1em}Transfer Attack & \textbf{0.1800} & \textbf{20.515} & \textbf{0.4445} & \textbf{0.3899} \\
\hspace{1em}Improv. (\%) & 42.4\% & 24.3\% & 36.5\% & 22.1\% \\
\midrule
\textbf{IMDB} \\
\hspace{1em}Direct Attack & 0.1137 & 34.805 & 0.2891 & \textbf{0.3923} \\
\hspace{1em}Transfer Attack & \textbf{0.1685} & \textbf{27.819} & \textbf{0.4333} & 0.3747 \\
\hspace{1em}Improv. (\%) & 48.1\% & 20.1\% & 49.8\% & -4.4\% \\
\midrule
\textbf{AGNEWS} \\
\hspace{1em}Direct Attack & 0.0538 & 286.16 & 0.1317 & 0.2742  \\
\hspace{1em}Transfer Attack & \textbf{0.0984} & \textbf{103.40} & \textbf{0.3589} & \textbf{0.3497} \\
\hspace{1em}Improv. (\%) & 82.9\% & 63.8\% & 172.5\% & 27.5\% \\
\bottomrule
\end{tabular}
}
\label{tab:main_result_different_domain}
\vspace{-3mm}
\end{table}
\label{sec:out_of_domain_attack}
To more comprehensively evaluate the capabilities of our methodology, we extended our evaluation of the transfer attack by incorporating an out-of-domain dataset, PersonaChat, as the external dataset, and present the result using SBERT as the victim embedding model in Table~\ref{tab:main_result_different_domain}. We have the following findings.
First, we found that utilizing an out-of-domain dataset is still helpful in improving attack performance. As shown in Table~\ref{tab:main_different_result_appendix}, transfer attack outperforms direct attack by roughly 20\%-40\% in QNLI and IMDB datasets.
Second, we notice that attacking with an out-of-domain dataset can achieve similar performance as the in-domain dataset. Specifically, the relative performance drop when utilizing an out-of-domain dataset is only 9.9\%, 3.1\%, and 14.5\% in embedding similarity and even lower in RougeL. This indicates that knowledge of the source domain is not always necessary for the attacker.
Due to the page limit, the full attack result is presented in Appendix~\ref{sec:main_diff_appendix}.

\section{Discussion}
\label{sec:surrogate-analysis}


\begin{table}[t!] 
\caption{Ablation study on the QNLI dataset. Rows shaded in grey represent results obtained using the Direct Attack method, while rows shaded in blue indicate the use of attack methods employing only surrogate models without additional training techniques.}

\label{tab:ablation}
\centering
\resizebox{\linewidth}{!}{
\begin{tabular}{ccccccc}
\toprule
\# $D_L$ & Surrogate & Adv. & Consist Reg. & RougeL & Cos & LLM-Eval   \\ 
\midrule
\rowcolor{gray!30}
\cellcolor{white} \multirow{5}{*}{500} & \xmark & \xmark & \xmark  & 0.0617  & 0.0609  & 0.2436 \\
\rowcolor{cyan!20}
 \cellcolor{white}& \cmark & \xmark & \xmark  & 0.1001  & 0.1310  & 0.2443 \\ 
 & \cmark & \cmark & \xmark  & 0.1192  & 0.1664  & 0.2550 \\ 
 & \cmark & \xmark & \cmark & 0.1251  & 0.1801  & \textbf{0.2686}  \\ 
 & \cmark & \cmark & \cmark & \textbf{0.1372}  &\textbf{ 0.2031}  & 0.2663  \\ 
\midrule
\rowcolor{gray!30}
\cellcolor{white} \multirow{5}{*}{8000} & \xmark & \xmark & \xmark  & 0.1264  & 0.3257  & 0.3194 \\
\rowcolor{cyan!20}
 \cellcolor{white} & \cmark & \xmark & \xmark  & 0.1701  & 0.4072  & 0.3598 \\ 
 & \cmark & \cmark & \xmark  & 0.1909  & 0.4742  & 0.4161 \\ 
 & \cmark & \xmark & \cmark & \textbf{0.1982}  & \textbf{0.4902}  & 0.4266  \\ 
 & \cmark & \cmark & \cmark & 0.1934  & 0.4886  & \textbf{0.4280}  \\  
\bottomrule 
\end{tabular}%
}
\end{table}





\subsection{Ablation Study}
Table~\ref{tab:ablation} is presented to study the effectiveness of each component. Here we consider three primary components: surrogate model, adversarial training, and consistency regularization.
Note that when all components are eliminated, our method becomes the direct attack method. Since the surrogate model is the key component of our transfer attack, we also highlight the performance of utilizing the surrogate model without any adjustment in blue. We first notice that using a surrogate model without training still improves the performance compared to direct attack, which indicates including additional training data could be helpful to some extent. Moreover, the surrogate model becomes better when either training objective is involved. For instance, the embedding similarity increases from 40\% to 47\% when including adversarial training when $|D_L$| is 8000. Utilizing consistency regularization could further boost the embedding similarity from 40\% to 49\%. Finally, the full model with all components could usually achieve the best or second-best performance, although we see diminishing returns as $|D_L|$ increases.


\begin{figure}[t!]
  \centering
  \includegraphics[width=0.9\linewidth]{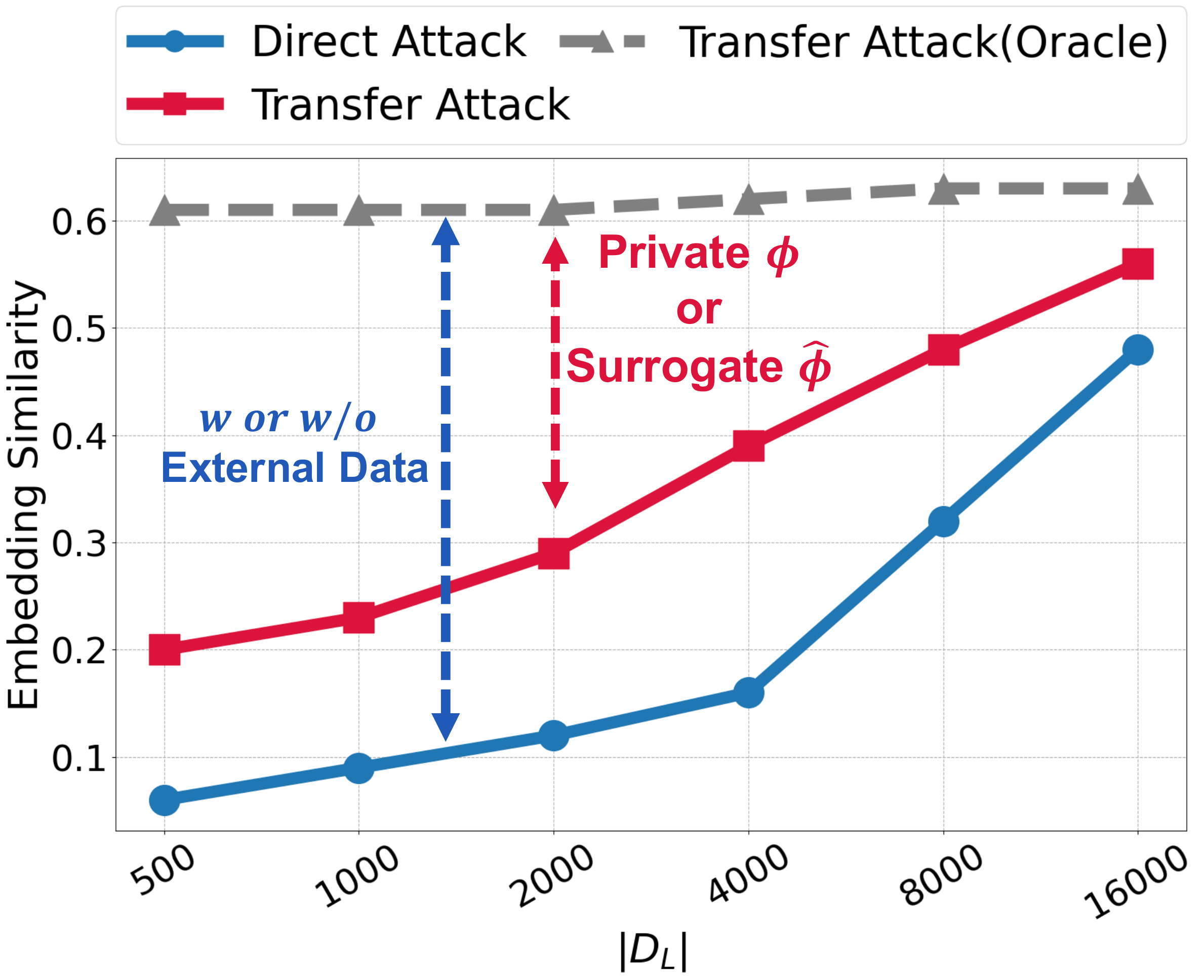}
  \caption{Comparison of attack performance on QNLI dataset $w.r.t.$ the amount of leaked dataset $D_L$.}
  \label{fig:train_size}
  \vspace{-6mm}
\end{figure}
\subsection{Size of the Leaked Dataset}
To understand the effectiveness of the surrogate model, we vary the number of leaked datasets to answer the following research questions. 

\noindent \textbf{How well does the surrogate model steal?} Although we have seen an improvement in transfer attack over direct attack, it still remains unclear to what extent the surrogate model steals the victim model $\phi$. Therefore, we implement the transfer attack(oracle) method which replaces the surrogate embedding with the actual embedding from the victim model.
The result is presented in Figure~\ref{fig:train_size}. Comparing direct attack and transfer attack(oracle), it is evident that utilizing external data could enhance the performance and thus a good surrogate model becomes essential for a successful attack. Generally, a more leaked dataset makes the surrogate model steal better and reaches its upper bound (i.e., the oracle model) when $|D_L|$ is 16000. Under our default setting where $|D_L|$ is 8000, the transfer attack achieves a score of 0.48, which is roughly 77\% of the upper limit's efficiency. Moreover, we also notice that the transfer attack is still effective when $|D_L|$ is small compared to a direct attack.

\begin{figure}[t!]
  \centering
  \includegraphics[width=0.9\linewidth]{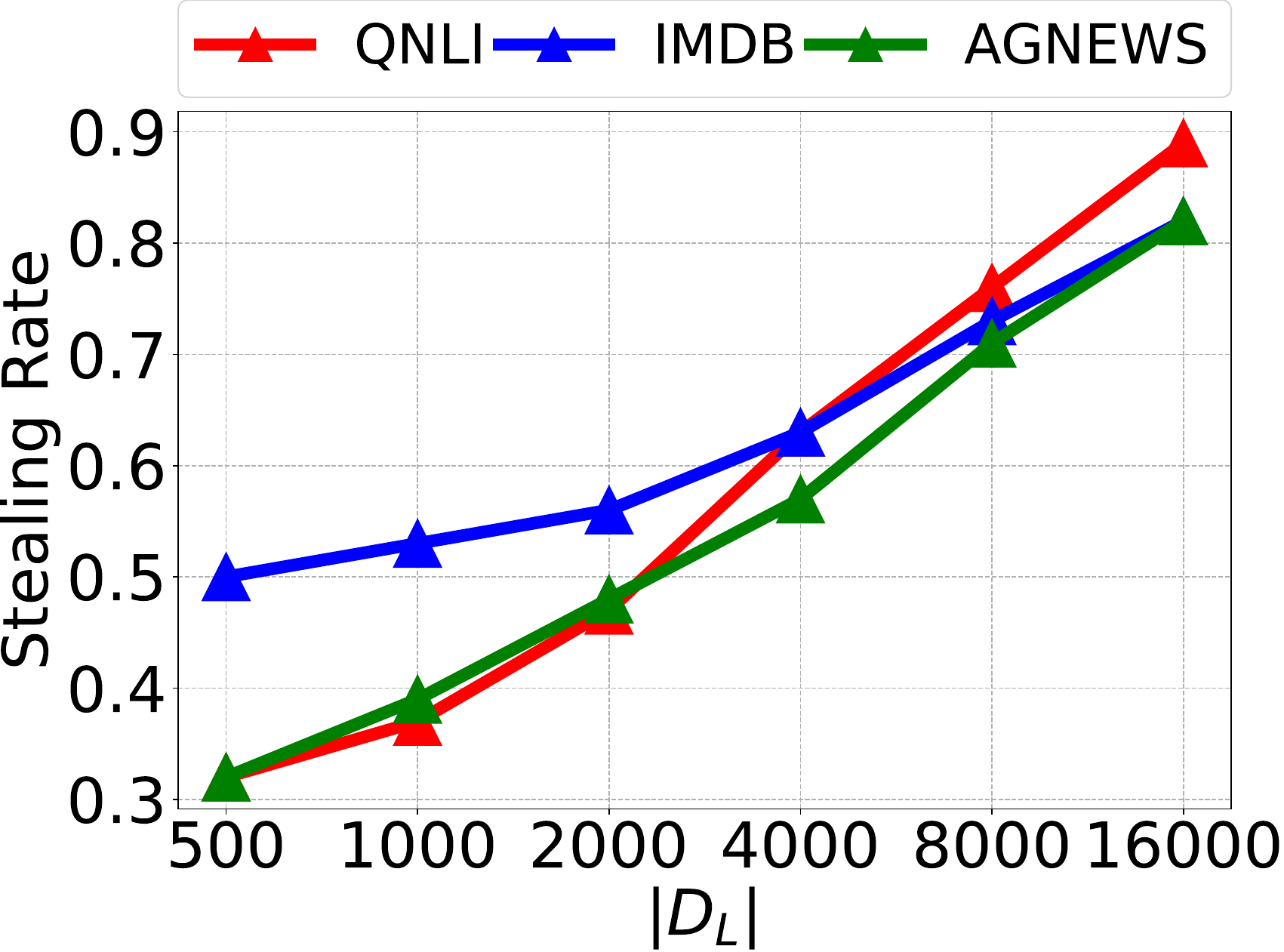}
  \caption{Stealing rate of the surrogate model compared to oracle model by varying the size of $D_L$.}
  \label{fig:stealing_rate}
\end{figure}

\noindent \textbf{How much data is the surrogate model required to be effective?}
To understand when the surrogate model can perform a successful steal $w.r.t.$ the amount of leaked data, we calculate the surrogate stealing rate by the ratio of attack performance with the transfer attack and the oracle model. In Figure~\ref{fig:stealing_rate}, the stealing rate across different datasets shows a similar trend. In general, the stealing rate achieves approximately 50\% when $|D_L|$ is 2000 and exceeds 70\% when $|D_L|$ is 8000. These results indicate our surrogate can effectively mimic the black box encoder with sufficient leaked data and reveal the privacy associated with the leaked data.

\subsection{Choice of a Surrogate Embedding Model}
\begin{figure}[t!]
  \centering
  \includegraphics[width=0.9\linewidth]{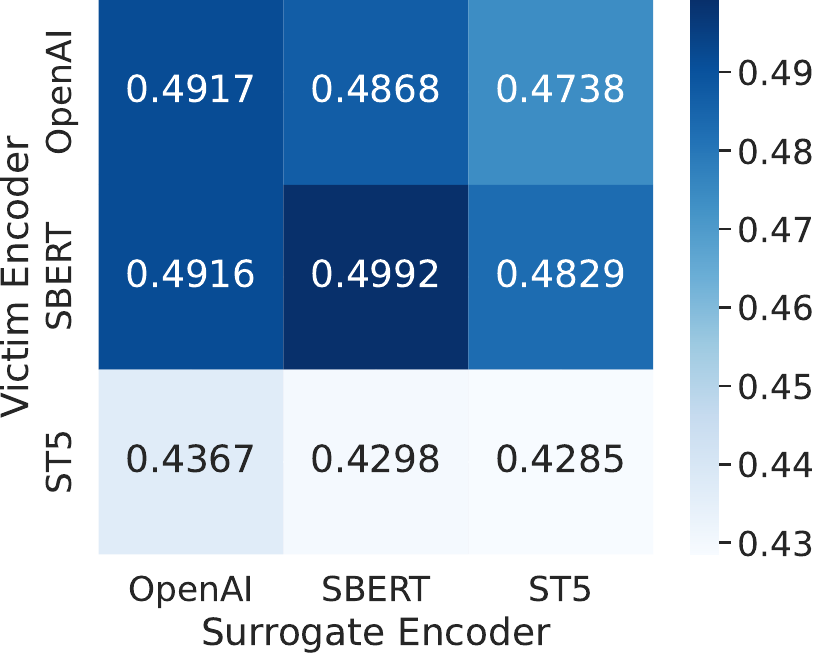}
  \caption{Attack performance by varying different victim and surrogate encoder. Here we use the embedding similarity metric to denote the attack performance.}
  \label{fig:surrogate}
\end{figure}

In this section, we explore how much the selection of a surrogate encoder affects the attack performance. Specifically, we use different surrogate encoders to attack embeddings generated with different victim encoders. The result is illustrated in Figure~\ref{fig:surrogate}. Next, we discuss the result in two aspects. 
\noindent \textbf{1) Is it necessary to know the target victim encoder?}
As the surrogate model is intended to replicate the behavior of the victim model, we seek to determine if employing an identical encoder enhances attack performance. 
The result of using an identical encoder can be found in the diagonal part of Figure~\ref{fig:surrogate}. Comparing the diagonal and non-diagonal parts, we see that using identical encoder attacks slightly better than those with different encoders in the case of OpenAI and SBERT. 
Moreover, when using ST5 as the victim model, selecting OpenAI or SBERT can even attack better than ST5.
This observation suggests that with our method, prior knowledge of the specific victim encoder is not required for a successful attack.
\noindent \textbf{2) Is our attack sensitive to the choice of the surrogate encoder?} In general, Figure~\ref{fig:surrogate} suggests that the attack performance does not vary too much when fixing a victim model. Specifically, the largest performance difference is 1.79\% for OpenAI, 1.63\% for SBERT, and 0.82\% for ST5. The result indicates that our attack pipeline is insensitive to the selection of the surrogate encoder.

\begin{table*}[t!]
\centering
\caption{Case study on the MIMIC-III dataset. We highlight the named entities (e.g., age, sex, disease, symptom and medical history) extracted by the biomedical NER model~\cite{biomedical-ner} for visualization.
}

\resizebox{\linewidth}{!}{
\begin{tabular}{lccc}
\toprule
Attack Methods & Sentences \\
\midrule
\textbf{Example 1} \\
\hspace{1em}Ground truth & \textcolor{red}{59 year-old} \textcolor{blue}{male} with a history of \textcolor{violet}{cardiomyopathy ef 45-50\% with pcm/icd} who presented due to \textcolor{orange}{sob}. \\
\hspace{1em}Transfer Attack & \textcolor{red}{59 year-old} \textcolor{blue}{male} with a past of \textcolor{violet}{cardiomyopathy ef 45-50\% with pcm/icd} who presented due to \textcolor{orange}{sob}. \\
\hspace{1em}Direct Attack & this is a 64 year old \textcolor{blue}{male} with known mitral regurgitation since.  \\
\midrule
\textbf{Example 2} \\
\hspace{1em}Ground truth & this is a \textcolor{red}{78 year-old} \textcolor{blue}{female} with a history of \textcolor{violet}{ild} who presents with \textcolor{teal}{altered mental status}.  \\
\hspace{1em}Transfer Attack & This is a \textcolor{red}{78 year-old} \textcolor{blue}{woman} with a history of \textcolor{violet}{ild} who presents with \textcolor{teal}{different mental status}.  \\
\hspace{1em}Direct Attack & this is an 80-year-old \textcolor{blue}{female} with a history of tracheobronchomalacia, copd, who presents with abdominal pain.  \\
\midrule
\textbf{Example 3} \\
\hspace{1em}Ground truth & this \textcolor{red}{73 year old} white \textcolor{blue}{male} has known \textcolor{orange}{aortic stenosis} which has progressed with increasing \textcolor{teal}{dyspnea}. \\
\hspace{1em}Transfer Attack & This \textcolor{red}{73 year old} white \textcolor{blue}{male} has identified \textcolor{orange}{aortic stenosis} which has progressed with worsening \textcolor{teal}{dyspnea}.  \\
\hspace{1em}Direct Attack & 67 year old \textcolor{blue}{male} with history of \textcolor{orange}{aortic stenosis} followed by serial echocardiograms.  \\
\bottomrule
\end{tabular}
}
\label{tab:case_study}
\end{table*}
\begin{table}[t!]
\centering
\caption{Embedding inversion performance evaluated with named entity recovery rate on MIMIC dataset. }
\resizebox{\linewidth}{!}{
\begin{tabular}{ccccccc}
\toprule
Attack Methods & Age & Sex & Disease & Symptom & History \\
\midrule
Transfer Attack & 98.84\% & 99.47\% & 79.07\% & 79.45\% & 65.36\% \\
Direct Attack & 7.79\% & 94.73\% & 19.35\% & 22.22\% & 17.49\% \\
\bottomrule
\end{tabular}
}
\label{tab:name_entity}
\end{table}

\section{Case Study}
\label{sec:case-study}
\subsection{Embedding inversion on MIMIC dataset.}
To demonstrate the privacy risks in a specific threat domain, we conduct a case study on MIMIC-III clinical notes~\cite{mimic}. The MIMIC-III dataset is a de-identified electronic health record database comprising comprehensive clinical data from intensive care units. Each note is truncated to its first sentence to remove redundant information. To be more realistic, 4K documents are sampled as the leaked dataset $D_L$. Following the out-of-domain attack setting, we choose PersonaChat as the external dataset.

To show the effectiveness of our transfer attack, we present the inverted result in Table~\ref{tab:case_study} using SBERT as the victim embedding model. For visualization, we highlight the named entities in each sentence as an indicator of sensitive attributes.
Comparing the inverted sentences, it is apparent that the transfer attack can almost recover the whole ground truth text, and the sensitive attributes are identified with high accuracy. However, the direct attack performs poorly due to the limited amount of leaked dataset.

\subsection{Recovery Rate on Named Entities}
To better understand how well can the transfer attack recover sensitive information from embeddings, we evaluate the named entity recover rate (NERR) and present the result in Table~\ref{tab:name_entity}. Specifically, we use the biomedical NER model~\cite{biomedical-ner} to extract named entities for each clinical note. The result exhibits that a transfer attack is able to recover 98\% of age and 99\% of sex. In particular, the transfer attack also achieves reasonable accuracy on disease, symptoms, and patient history and outperforms direct attack with a significant improvement. In summary, we found that the transfer attack can indeed reveal more privacy risks than a standard attack method.
\section{Related Work}

\textbf{Inversion attacks on embeddings.}
Embedding inversion attacks have been explored across computer vision~\cite{bordes2022high, dosovitskiy2016inverting, teterwak2021understanding} and NLP~\cite{privacy_risks} domains with significant implications for privacy. Typically, these inversion attacks make assumptions about the attacker's access to the victim model and evaluate the associated privacy risks.
White-box scenarios assume attacker access to the full model weights, this enables the attack to approximate the inverse function with nearly 100\% recover rate of text sequences~\cite{kugler2021invbert}. Existing black-box attacks~\cite{privacy_risks, GEIA} assume an attacker has no knowledge of the underlying model itself, and can only interact with models with the query access. A recent work~\cite{morris2023text} demonstrated an iterative recovery process that can reconstruct 92\% of a 32-token text. 
Existing embedding inversion research largely depends on querying the victim model, yet the unexplored potential of query-free attacks presents a valuable opportunity for the community.

\noindent \textbf{Stealing attacks on ML models.}
Many works in model stealing focus on stealing classifiers. In general, these methods steal the exact model parameters or functionality of target classifiers by querying them. For instance, prior works studied stealing ML models (e.g., decision tree or neural networks) deployed on cloud services. In a similar line, a few recent research~\cite{cong2022sslguard,liu2022stolenencoder} proposed an attack to steal a pre-trained encoder. By stealing the encoder, the attacker can obtain similar functionality on downstream tasks. For instance, StolenEncoder~\cite{liu2022stolenencoder} demonstrated the effectiveness of stealing powerful encoders (e.g., CLIP~\cite{radford2021learning} by OpenAI) with a ResNet-34 model. Similarly, there are several research~\cite{naseh2023stealing,zanella2021grey} on stealing language models. Specifically, attacks on BERT-based APIs~\cite{krishna2020thieves,he2021model} show that attackers can steal effectively via querying it without knowing the training data of the target language model. Different from the prior stealing attacks which steal encoders for downstream applications, our work leverages the stolen encoder to facilitate transfer attacks on the target encoder.
\section{Conclusion}
In this work, we study the privacy risks associated with text embeddings, especially under constraints where attackers lack direct query access to the embedding models. Through the development of a transfer attack method, we demonstrated the feasibility of inferring sensitive information from embeddings without needing to interact with the original model. Our extensive experiments across various embedding models and a detailed case study on a clinical dataset underline the effectiveness of our approach. As the use of text embeddings continues to grow in a wide range of applications, our work serves as a crucial step toward understanding and mitigating potential privacy threats.

\newpage
\section{Limitations}
The primary limitation of our attack methodology is its ineffectiveness in accurately reconstructing longer sentences. Notably, when evaluating the effectiveness of our transfer attack on the AG News dataset, as depicted in Table~\ref{tab:main_result_same_domain}, we observe that the RougeL scores are $5\%$ to $7\%$ lower than those achieved on the IMDB dataset, regardless of the black box embedding algorithm employed. Additionally, the model demonstrates the highest Perplexity (PPL) score on the AG News dataset in comparison to others, indicating a notable instability in model predictions. Table~\ref{tab:dataset} further highlights that AG News has the longest average sentence length among the datasets examined. This trend suggests that our transfer attack approach encounters difficulties in the effective reconstruction of longer sentences. 

\section{Acknowledgement}
This material is based upon work supported by National Science and Technology Council, ROC under grant number 111-2221-E-002 -146 -MY3 and 112-2634-F-002 -005

\bibliography{acl_latex}
\bibliographystyle{acl_natbib}

\appendix

\begin{table}[ht!]
\centering
\caption{Statistics of datasets.}
\resizebox{\linewidth}{!}{
\begin{tabular}{lcccc}
\toprule
Statistic Type & QNLI & IMDB & AGNEWS & PersonaChat \\
\midrule
Task & NLI & Sentiment & Classification & Dialog \\
Domain & Wikipedia & Reviews & News & Chit-chat \\
Avg. sent length & 18.25 & 21.14 & 28.09 & 10.12 \\ 
Unique words & 799519 & 1031895 & 1372484 & 1639738 \\
\bottomrule
\end{tabular}
}
\label{tab:dataset}
\end{table}
\section{Detailed Dataset Statistics}
\label{appendix:dataset-stats}
Table~\ref{tab:dataset} compares four datasets—QNLI, IMDB, AGNEWS, and PersonaChat—across various metrics. QNLI, focusing on Natural Language Inference from Wikipedia, has an average sentence length of 18.25 words and 799,519 unique words. IMDB, for sentiment analysis from movie reviews, has an average sentence length of 21.14 words and 1,031,895 unique words. AGNEWS, aimed at news classification, features the longest sentences on average (28.09 words) and 1,372,484 unique words. PersonaChat, designed for dialog in chit-chat, has the shortest sentences (10.12 words) but the largest vocabulary, with 1,639,738 unique words. This summary showcases the datasets' diversity in application, domain, and linguistic characteristics.

\section{Hyperparameters}
We utilized pretrained DialoGPT-small~\cite{Dialogpt} as our specified attack model, while employing GTR-base~\cite{GTR} as our surrogate encoder. For optimization, we employed the AdamW~\cite{AdamW} optimizer with a learning rate of $3\times10^{-5}$ alongside warmup and linear decay, using a batch size of 16.Under these conditions, our model undergoes training for approximately 15 hours.

\section{Full Out-of-Domain Experiment}
\label{sec:main_diff_appendix}
\begin{table*}[t!]
\centering
\caption{Comparison of different domain embedding inversion performance between direct and transfer attack. The evaluation are done on QNLI, IMDB and AGNEWS datasets with embedding models including: OpenAI \texttt{text-embeddings-ada-002}, SBERT~\cite{sentenceBERT} and ST5~\cite{sentence-T5}.}
\resizebox{\textwidth}{!}{
\begin{tabular}{@{}lcccccccccccc@{}}
\toprule
\multirow{2}{*}{Dataset / Method} & \multicolumn{4}{c}{OpenAI} & \multicolumn{4}{c}{SBERT} & \multicolumn{4}{c}{ST5} \\
\cmidrule(lr){2-5} \cmidrule(lr){6-9} \cmidrule(lr){10-13}
& RougeL & PPL & Cos & LLM-Eval & RougeL & PPL & Cos & LLM-Eval & RougeL & PPL & Cos & LLM-Eval \\
\midrule
\textbf{QNLI} \\
\hspace{1em}Direct Attack & 0.1433 & 40.822 & 0.2797 & 0.2984 & 0.1264 & 27.127 & 0.3257 & 0.3194 & 0.1463 & 42.911 & 0.2226 & 0.2755 \\
\hspace{1em}Transfer Attack & \textbf{0.2071} & \textbf{18.692} & \textbf{0.4253} & \textbf{0.3987} & \textbf{0.1800} & \textbf{20.515} & \textbf{0.4445} & \textbf{0.3899} & \textbf{0.1931} & \textbf{19.829} & \textbf{0.3946} & \textbf{0.3825} \\
\hspace{1em}Improv. (\%) & 44.5\% & 54.2\% & 52.0\% & 33.6\% & 42.4\% & 24.3\% & 36.5\% & 22.1\% & 31.9\% & 53.8\% & 77.2\% & 38.8\% \\
\midrule
\textbf{IMDB} \\
\hspace{1em}Direct Attack & 0.1133 & \textbf{20.549} & 0.2692 & 0.3818 & 0.1137 & 34.805 & 0.2891 & \textbf{0.3923} & 0.1103 & \textbf{24.939} & 0.2678 & 0.3909 \\
\hspace{1em}Transfer Attack & \textbf{0.1808} & 25.756 & \textbf{0.4157} & \textbf{0.4504} & \textbf{0.1685} & \textbf{27.819} & \textbf{0.4333} & 0.3747 & \textbf{0.1563} & 30.311 & \textbf{0.3792} & \textbf{0.4398} \\
\hspace{1em}Improv. (\%) & 59.6\% & -25.3\% & 54.4\% & 17.9\% & 48.1\% & 20.1\% & 49.8\% & -4.4\% & 41.7\% & -21.5\% & 41.6\% & 12.5\% \\
\midrule
\textbf{AGNEWS} \\
\hspace{1em}Direct Attack & 0.0612 & \textbf{66.383} & 0.1162 & 0.2979 & 0.0538 & 286.16 & 0.1317 & 0.2742 & 0.0578 & \textbf{74.085} & 0.0980 & 0.2905 \\
\hspace{1em}Transfer Attack & \textbf{0.1066} & 101.04 & \textbf{0.3655} & \textbf{0.3618} & \textbf{0.0984} & \textbf{103.40} & \textbf{0.3589} & \textbf{0.3497} & \textbf{0.0938} & 128.26 & \textbf{0.3256} & \textbf{0.3460} \\
\hspace{1em}Improv. (\%) & 74.1\% & -52.2\% & 214.5\% & 21.4\% & 82.9\% & 63.8\% & 172.5\% & 27.5\% & 62.2\% & -73.1\% & 232.2\% & 19.1\% \\
\bottomrule
\end{tabular}
}
\label{tab:main_different_result_appendix}
\end{table*}

The detailed results for different domain experiment are presented in Table~\ref{tab:main_different_result_appendix}. In the majority of scenarios, our approach surpasses the baseline method across several evaluation metrics, with the exception of perplexity. The data indicates an improvement exceeding $40\%$ in embedding similarity scores and a $30\%$ enhancement in RougeL scores.

\section{Comparison of Augmentation Strategies}
\label{appendix:augmentation}

Upon reviewing the results presented in Table~\ref{tab:different_augmentation_result}, it is evident that Large Language Model-based (LLMDA) approaches exhibit superior performance, consistently ranking either as the best or second-best across all metrics. Notwithstanding, notable observations merit attention: RougeL and Cosine Similarity  metrics for the Swap approach surpass those of LLM. This discrepancy can be attributed to RougeL and Cosine Similarity placing lesser emphasis on the sequential order of generated sentences. Additionally, the Swap method avoids introducing out-of-vocabulary words, a characteristic of potential significance, whereas LLM may generate words not present in the original dataset. These considerations contribute to the observed outcome wherein Swap outperforms LLM with respect to RougeL and Cosine Similarity metrics. However, a nuanced examination of specific sentences generated by LLM and Swap reveals the discernible superiority of sentences produced by LLM.
\begin{table}[t!]
\centering
\caption{Comparison of embedding inversion performance between different data augmentation approaches. We bold the best performance and underline the second-best performance in the table.}
\resizebox{\linewidth}{!}{
\begin{tabular}{ccccc}
\toprule
Method & RougeL & PPL & Cos & LLM-Eval \\
\midrule
LLM & \underline{0.1782} & \textbf{18.062} & \underline{0.4496} & \underline{0.3976} \\
Swap & \textbf{0.1932} & 35.587 & \textbf{0.5488} & 0.3764 \\
Delete & 0.1579 & 19.944 & 0.3950 & \textbf{0.4150} \\
Replace & 0.1727 & 24.991 & 0.4120 & 0.3393 \\
Insert & 0.1138 & \underline{18.644} & 0.3225 & 0.2387 \\
\rowcolor{Gray}
w/o Aug. & 0.1490 & 23.237 & 0.3419 & 0.3270 \\
\bottomrule
\end{tabular}
}
\label{tab:different_augmentation_result}
\end{table}

\section{Prompts for LLM Data Augmentation}
We offer the following prompt for LLM Data augmentation, with the constraint of 2 words explicitly within the prompt. \\
\\
\underline{\textbf{Prompt template}}: \\
\\
{\color{gray}Please rewrite the original sentence with synonyms within 2 words. \\
Please output 5 different new sentences. \\
Please simply modify the original sentence without changing more than 2 words. \\
\\
Example: \\
Original sentence: \\
\{ORIGINAL SENTENCE\} \\
New sentence: \\
\{NEW SENTENCE 1\} \\
\{NEW SENTENCE 2\} \\
\{NEW SENTENCE 3\} \\
\{NEW SENTENCE 4\} \\
\{NEW SENTENCE 5\} \\
\\
Original sentence: \\
\{INPUT SENTENCE\}  \\
New sentence:}

\section{Details of LLM Evaluation}
\label{appendix:llm-eval}
We assess the outcome using a large language model to closely emulate human assessment. Our evaluation metric aims to gauge semantic similarity, fluency, and coherence between the prediction and the ground truth sentence. Below is the prompt template utilized for this purpose. \\
\\
\underline{\textbf{Input prompt}}: \\
\\
{\color{gray}Output a number between 0 and 1 describing the semantic similiarity, fluent ,and coherent between the following two sentences: please output the answer without any explaination. \\
\{pred sentence\} \\
\{ground truth sentence\}} \\

\section{More Case Study}
Table~\ref{tab:case_study_app} presents another case study conducted on the QNLI dataset, utilizing SBERT as the target embedding model. To aid visualization, we highlight the informative words within the ground truth sentences. Where inverted sentences contain sensitive named entities with analogous meanings, we have applied corresponding color highlights. This outcome underscores the effectiveness of transfer attacks in accurately recovering informative words, whereas direct attacks often result in erroneous accompanying information in the majority of cases.
\begin{table*}[t!]
\centering
\caption{Case study on QNLI dataset. In the ground truth sentence, place is represented by red, time by purple, other noun by blue, verb by orange, and adjective by green.}
\resizebox{\linewidth}{!}{
\begin{tabular}{lcc}
\toprule
Attack Methods & Sentence \\
\midrule
\textbf{Example 1} \\
\hspace{1em}Ground truth & Who \textcolor{orange}{founded} the \textcolor{blue}{city} of \textcolor{red}{London}? \\
\hspace{1em}Transfer Attack & Who \textcolor{orange}{founded} the \textcolor{blue}{city} of \textcolor{red}{London}? \\
\hspace{1em}Direct Attack & Which county in the Anglo-Saxon Empire? \\
\midrule
\textbf{Example 2} \\
\hspace{1em}Ground truth & What is the \textcolor{teal}{largest} \textcolor{blue}{bird}? \\
\hspace{1em}Transfer Attack & What is the \textcolor{teal}{most largest} \textcolor{blue}{bird}? \\
\hspace{1em}Direct Attack & How many animals inhabit the Tuna beak are various plankton Empire? \\
\midrule
\textbf{Example 3} \\
\hspace{1em}Ground truth & What was \textcolor{red}{Nigeria's} \textcolor{blue}{population} in \textcolor{violet}{2009}? \\
\hspace{1em}Transfer Attack & What was \textcolor{red}{Nigeria's} \textcolor{blue}{population} in 2011? \\
\hspace{1em}Direct Attack & What was the \textcolor{blue}{total number of people} in the Middle East who had Internet before 2010? \\
\midrule
\textbf{Example 4} \\
\hspace{1em}Ground truth & What \textcolor{blue}{Air Force base} is in \textcolor{red}{Tucson}? \\
\hspace{1em}Transfer Attack & What \textcolor{blue}{military} airport is in \textcolor{red}{Tucson}? \\
\hspace{1em}Direct Attack & What is the total land area of the \textcolor{blue}{Army base} on the Eisenhower Parkway? \\
\midrule
\textbf{Example 5} \\
\hspace{1em}Ground truth & Who \textcolor{orange}{established} the \textcolor{red}{Tibetan} \textcolor{blue}{law code}? \\
\hspace{1em}Transfer Attack & Who implemented the \textcolor{red}{Tibetan} \textcolor{blue}{Penal Code}? \\
\hspace{1em}Direct Attack & How was the \textcolor{red}{Tibetan} Buddhists' policy on the TB inconsistent with secular practices? \\
\bottomrule
\end{tabular}
}
\label{tab:case_study_app}
\end{table*}




\end{document}